\documentclass[sigconf]{acmart}
\usepackage{multirow}
\usepackage{tabularx}
\usepackage{subfigure}
\usepackage{cleveref}
\usepackage{hyperref}
\usepackage{graphicx}

\newcolumntype{L}{>{\RaggedRight\hangafter=1\hangindent=0em}X}

\AtBeginDocument{%
  }

\setcopyright{acmlicensed}
\copyrightyear{2018}
\acmYear{2018}
\acmDOI{XXXXXXX.XXXXXXX}

\acmConference[Conference acronym 'XX]{Make sure to enter the correct
  conference title from your rights confirmation emai}{June 03--05,
  2018}{Woodstock, NY}
\acmISBN{978-1-4503-XXXX-X/18/06}

\begin{document}

\title{HACD: Harnessing Attribute Semantics and Mesoscopic Structure for Community Detection}

\author{Anran Zhang}
\email{2022020934@bistu.edu.cn}
\orcid{0009-0005-5350-0281}
\affiliation{%
  \institution{Beijing Information Science and Technology University}
  \city{Beijing}
  \country{China}
}

\author{Xingfen Wang}
\authornote{Corresponding author.}
\email{xfwang@bistu.edu.cn}
\affiliation{%
  \institution{Beijing Information Science and Technology University}
  \city{Beijing}
  \country{China}}

\author{Yuhan Zhao}
\email{csyhzhao@comp.hkbu.edu.hk}
\affiliation{%
  \institution{Hong Kong Baptist University}
  \city{Hong Kong}
  \country{China}
}

\renewcommand{\shortauthors}{Trovato et al.}

\begin{abstract}
Community detection plays a pivotal role in uncovering closely connected subgraphs, aiding various real-world applications such as recommendation systems and anomaly detection. With the surge of rich information available for entities in real-world networks, the community detection problem in attributed networks has attracted widespread attention. While previous research has effectively leveraged network topology and attribute information for attributed community detection, these methods overlook two critical issues: (i) the semantic similarity between node attributes within the community, and (ii) the inherent mesoscopic structure, which differs from the pairwise connections of the micro-structure. To address these limitations, we propose HACD, a novel attributed community detection model based on heterogeneous graph attention networks. HACD treats node attributes as another type of node, constructs attributed networks into heterogeneous graph structures and employs attribute-level attention mechanisms to capture semantic similarity. Furthermore, HACD introduces a community membership function to explore mesoscopic community structures, enhancing the robustness of detected communities. Extensive experiments demonstrate the effectiveness and efficiency of HACD, outperforming state-of-the-art methods in attributed community detection tasks. Our code is publicly available at \href{https://github.com/Anniran1/HACD1-wsdm}{https://github.com/Anniran1/HACD1-wsdm}.
\end{abstract}



\maketitle
\section{Introduction}
Community detection \cite{wu2022clare} is a fundamental problem in network analysis, seeking to unveil closely connected subgraphs (i.e., communities) within complex networks. 
Previous research has adeptly utilized network topology to discern communities \cite{kang2021adversarial,hou2022meta}. However, nodes in real-world networks typically possess rich attribute information. For example, in citation networks \cite{ye2023top}, papers are associated with specific keyword domains. Such networks, known as attributed graphs  \cite{yang2013community}, introduce additional complexity for community detection algorithms. 

To harness the potential of topology and attribute information for attributed community detection (ACD), existing methods, e.g., CommDGI \cite{zhang2020commdgi} and ACDM \cite{cheng2023significant}, map these dual information sources to low-dimensional continuous vector spaces by using embedding techniques. 
While these methods have demonstrated promising results, we contend that current solutions may not be optimal because they overlook two critical issues:
\begin{figure}
  \includegraphics[width=0.8\linewidth]{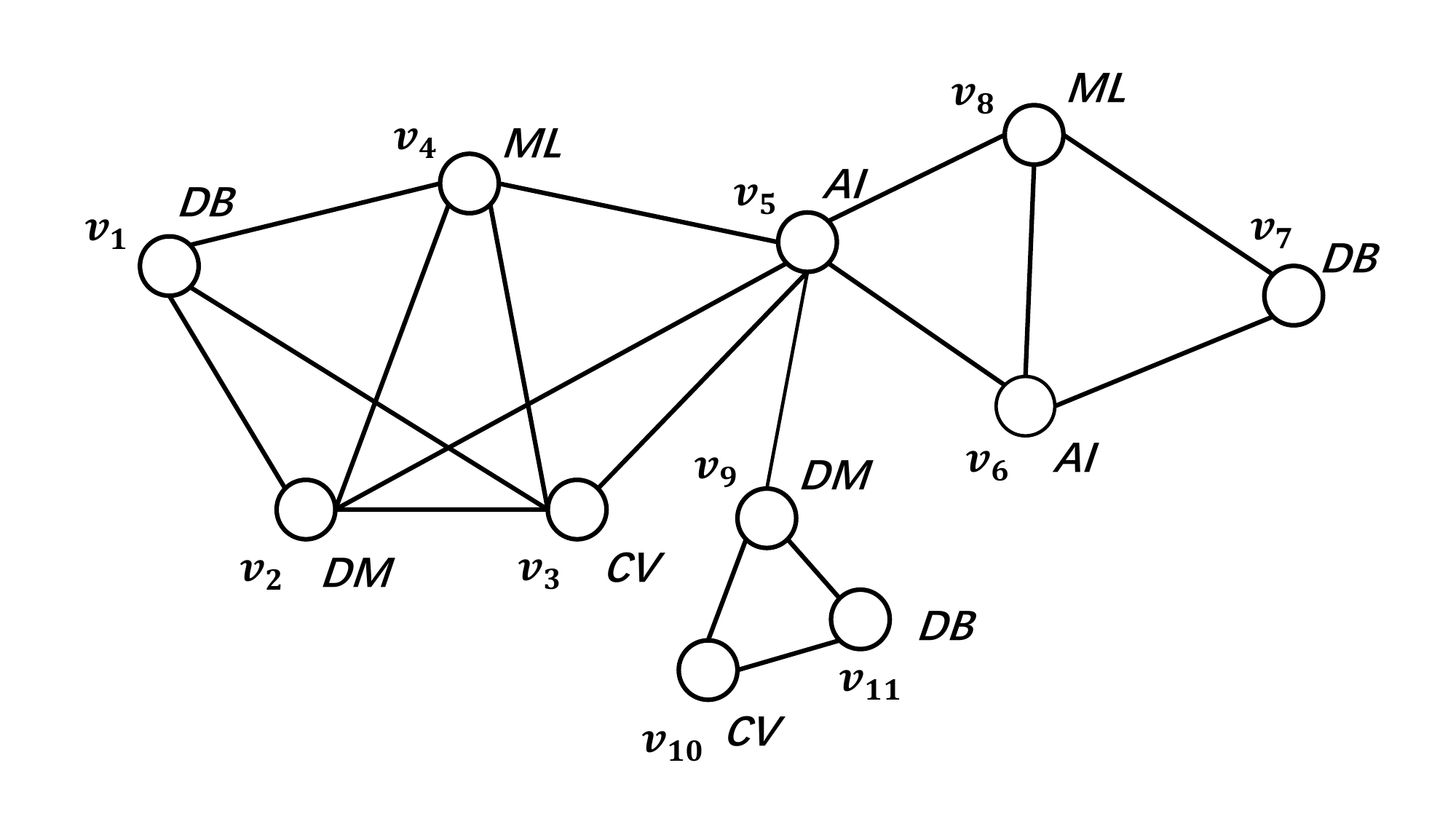}
  \caption{Most studies treat AI (artificial intelligence), CV (computer vision), and ML (machine learning) as independent attributes. However, AI and CV are subfields within the broader domain of ML, implying that they share underlying semantic similarities.}
  \label{fig:example}
\end{figure}
\begin{itemize}
    \item \textbf{Semantic similarity.} 
    Semantic similarity refers to the degree of semantic resemblance or the extent of correlation between attributes. For instance, as illustrated in Figure~\ref{fig:example}, the semantic similarity of attributes can reveal latent relationships between nodes and enhance the attribute cohesiveness of detected communities\cite{jiang2021query}.
    However, existing methods usually disregard the semantic similarity between node attributes within communities, leading to the omission of crucial nodes in the detected communities.
    \item \textbf{Mesoscopic community structure.} 
    Inherent community structure, serving as a crucial mesoscopic description of network topology, imposes constraints on node representation at a higher structural level. If the mesoscopic community structure is considered to guide network embedding, the results would remain robust against minor local changes in the network structure, such as node noise and the addition or deletion of edges or nodes \cite{liu2022robust}. However, existing methods primarily focus on the pairwise connections of micro-structure between nodes \cite{zhang2021spectral}, rendering the results overly sensitive to minor changes in microscopic structure.
\end{itemize}

To address these limitations, we propose a novel attributed community detection model based on a heterogeneous graph attention network (HAN), termed HACD. To tackle the first issue, we initially treat node attributes as another type of node, transforming real-world attributed networks into a heterogeneous graph structure. Subsequently, we propose an attribute-level attention mechanism (A2M), which utilizes weighted aggregation based on attention coefficients to identify key attributes within each community and employs an attention-based similarity metric to compute the distance between the semantic meanings of different attributes. By embedding with A2M, the representation learns the importance of different attributes and captures the semantic similarity between node attributes. This semantic similarity fully reflects the latent relationships between nodes, achieving attribute cohesion within communities. Furthermore, to address sensitivity issues and enhance robustness, we introduce a community membership function (CMF). By encoding initial community membership information and introducing a new modularity function to formulate CMF as a modularity optimization problem, we guide network embedding to explore mesoscopic community structures, ensuring the structural cohesiveness of detected communities. 

Our principal contributions can be summarized as follows:
\begin{itemize}
\item We first identify two critical problems affecting attributed community detection: semantic similarity and mesoscopic community structure. 
\item We propose a novel attributed community detection model, HACD. We construct the attribute network as a heterogeneous graph structure and introduce the heterogeneous graph neural network into attributed community detection tasks. We propose an attribute-level attention method to explore the semantic similarity between node attributes, as well as design a community membership function to obtain the mesoscopic community structure. 
\item We conduct extensive experiments demonstrating the effectiveness and efficiency of HACD, showing superiority over state-of-the-art community detection methods in attributed graph datasets.
\end{itemize}
\section{Preliminaries}
\subsection{Problem Statement}
\subsubsection{Attributed Network} 
An attributed network \cite{cen2019representation} is typically denoted by a graph $G=(V,E,A)$, where $V=\left\{v_1, v_2, \ldots, v_n\right\}$ represents the set of $n$ nodes. $E \subset\left\{\left(v_i, v_j\right) \mid v_i, v_j \in V\right\}$ is the edges sets where each edges connect two nodes in the graph. $A=\{ a_1, a_2, \ldots, a_n \}$ is the set of node attributes for all nodes, where $a_i$ is the attributes of node $v_i$. In addition, each node $v_i \in V$ is associated with some types of $d$-dimensional attribute feature vectors, the feature matrix can be represented as $\textbf{X}=\left\{\textbf{x}_1, \textbf{x}_2, \ldots, \textbf{x}_n\right\}^T \in \mathbb{R}^{n \times d}$.

\subsubsection{HACD-Problem} 
Given an attributed network $G=(V,E,A)$, the problem of attributed community detection based on heterogeneity returns attributed communities $C=\left\{c_1, c_2, \ldots, c_k\right\}$ from a heterostructure aspect, satisfying the following properties (i) \textbf{structure cohesiveness,} where nodes within each community are tightly connected, while nodes in different communities are sparsely connected, and (ii) 
\textbf{attribute cohesiveness,} where the attributes of nodes within a community have a semantic similarity.

\subsection{Attribute cohesiveness.}
Usually, the attribute score \cite{liu2020vac} is used to measure the attribute cohesiveness of a community. Given two nodes $u$, $v$, the attribute score is denoted as $Ascore(u,v)$. For different types of attributes, we can employ different methods, such as $Euclidean$ $distance$ and $Jaccard$ $distance$ \cite{kosub2019note}, to calculate the attribute score of two nodes. When different types of attributes co-exist, we can employ a unified function to combine different distance functions, e.g., $Ascore(u,v)=\alpha \cdot \frac{Sdist(u,v)}{Sdist_{max}} + (1-\alpha) \cdot \frac{Tdist(u,v)}{Tdist_{max}}$, where $Sdist(u,v)$ and $Tdist(u,v)$ compute the numerical distance and textual distance, respectively; $Sdist_{max}$ and $Tdist_{max}$ are the maximum numerical distance and maximum textual distance, respectively, for normalization; the parameter $0\leq \alpha \leq 1$ is used to balance numerical proximity and textual relevancy.

\subsection{Structure cohesiveness.}
Modularity reflects the quality of community structure in a network, which is a commonly used performance metric to measure the structure cohesiveness of communities\cite{rustamaji2024community,zhang2023detecting}. The traditional definition of modularity\cite{newman2006finding} is based on the adjacency matrix of a graph, and the modularity function is defined as follows:
\begin{equation}
Q=\frac{1}{2M} \sum_{i,j} (\textbf{A}_{ij} - \frac{k_i k_j}{2M})  \delta(c_i,c_j)
\end{equation} 
where $M$ denotes the number of edges in the graph; $\textbf{A}_{ij}$ can be understood as the observed structural information between two nodes $v_i$, $v_j$, for example, the edge between nodes $v_i$ and $v_j$; $k_i$ denotes the degree of node $v_i$; as well as $\delta(c_i,c_j)$ denotes the connectivity between community $c_i$ and community $c_j$, which can be calculated based on the community division. A common way to calculate it is to define a connectivity matrix $\delta$, where $\delta_ij$ denotes whether node $v_i$ and node $v_j$ are in the same community, i.e., $\delta(c_i,C_j)=1$ if $c_i=c_j$, 0 otherwise.
\begin{figure*}
  \includegraphics[width=0.8\textwidth]{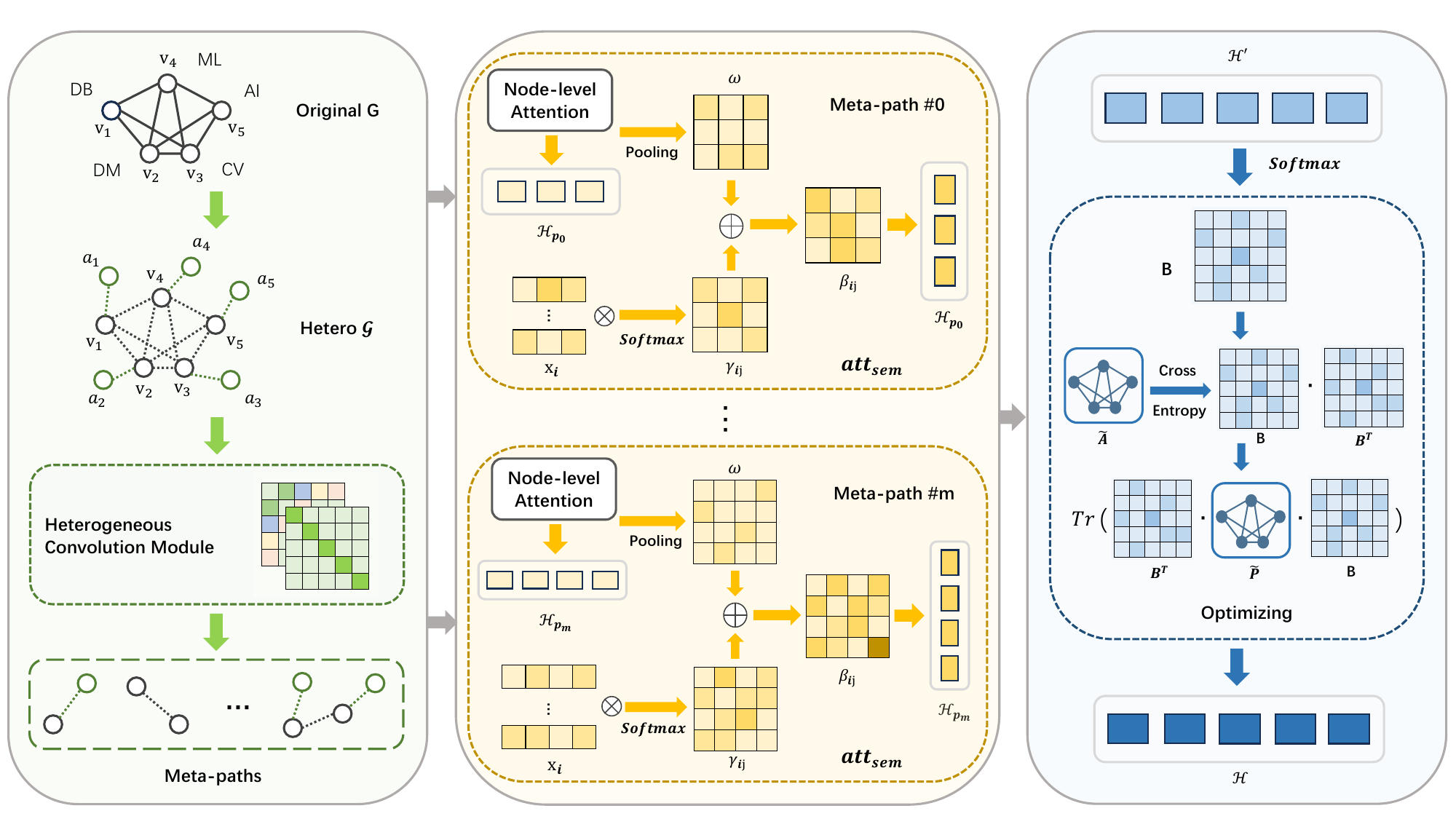}
  \caption{The overall framework of HACD. }
  \label{fig:framework}
\end{figure*}
\section{Methodology}

\subsection{Overall Framework}
To address the above challenges, we introduce a novel HACD method, as shown in Figure~\ref{fig:framework}, that comprises three key components: \textbf{graph constructing and encoding}, \textbf{attribute-level attention mechanism}, and \textbf{community membership function}.

\subsection{Graph Constructing and Encoding}
Traditional graph construction methods typically rely on homogeneous structures, where attribute information is directly encoded, making it challenging to uncover the semantic nuances therein. To delve deeper into attribute information, departing from conventional homogeneous attribute graph approaches, we introduce heterogeneous graphs, treating node attributes as an additional node type to catch semantic similarity. 

Our approach centers on the construction of a heterogeneous attribute graph, denoted as $G=(V,E,A)$. Initially, we define the node and edge types within the desired heterogeneous graph. Here, the original attribute information $A=\{a_1, a_2, \ldots, a_n \}$ is treated as an additional node type, complementing the intrinsic node types $V= \{v_1, v_2, \ldots, v_n \}$ within $G$. The resulting set of nodes in the heterogeneous graph is represented as $\mathcal{V}$, encompassing various distinct node types. The relationship between the original node entities and attribute node entities is delineated by possession, refining $E$ to $\mathcal{E} \subset \{(v_i, v_j),(v_i, a_i) \mid v_i, v_j \in V, a_i \in A \}$. Thus, we derive the heterogeneous graph $\mathcal{G}=(\mathcal{V},\mathcal{E})$. Notably, $\mathcal{G}$ is characterized by a node type mapping function $\phi(v): \mathcal{V} \rightarrow \mathcal{A}$ and an edge type mapping function $\phi(e): \mathcal{E} \rightarrow \mathcal{R}$, where $\mathcal{A}$ and $\mathcal{R}$ denote the sets of predefined node types and edge types, respectively, with $|\mathcal{A}|+|\mathcal{R}|>2$.

To address the challenge of heterogeneity, meta-paths have become a staple in various heterogeneous graph embedding methodologies. However, traditional approaches often rely on manually predefined meta-paths, necessitating expert prior knowledge and potentially impacting model efficacy. In this work, we introduce a novel heterogeneous convolution module \cite{chang2022megnn}, denoted as $F ^{(l)}(\cdot)$, designed to automatically generate and extract effective meta-path schemes. Mathematically, the module operates as follows:
\begin{equation}
A_{conv} ^{(l)}=F^{(l)}( { A_e | e \in \mathcal{T}^e }) = \sum_{e \in \mathcal{T}^e} \alpha_e A_e,
\end{equation}
where, $A_{conv} ^{(l)}$ and ${ A_e | e \in \mathcal{T}^e }$ represent the set of local bipartite graphs in $\mathcal{G}$ and the edge type set of the graph $\mathcal{T}^e$, respectively. $\alpha^e$ denotes a layer-wise independent parameter to be learned, signifying the contribution of the sub-graph of type $e$ to the convolved structures. Recognizing that the neighbors of each node play distinct roles and carry varying degrees of importance in learning node embeddings, we then propose to embed nodes using node-level attention inspired by HAN \cite{wang2019heterogeneous}, facilitating the capture of complex structures and rich semantic information. Specifically, we compute the importance between node pairs based on meta-paths and normalize them to obtain the weight coefficient $\alpha_{ij} ^p$ via the softmax function:
\begin{equation}
\alpha_{ij} ^p= \frac{exp(\sigma(a_p ^T \cdot [ \textbf{h}_i ^{\prime} || \textbf{h}_j ^{\prime}]))}{\sum_{k \in \mathcal{N}_i ^p}exp(\sigma(a_p ^T \cdot [ \textbf{h}_j ^{\prime}|| \textbf{h}_k ^{\prime}]))},
\end{equation} 
where $\sigma$ denotes the activation function, $||$ signifies the concatenation operation, $a_p$ represents the node-level attention vector for the meta-path, and $\textbf{h}_i ^{\prime}$ projects the features of different node types into the same feature space.

Subsequently, the meta-path-based embedding of node $i$ is aggregated by the projected features of its neighbors with the corresponding coefficients as follows \cite{wang2019heterogeneous}:
\begin{equation}
\textbf{h}_i ^p=\sigma(\sum_{j \in \mathcal{N}_i ^p} \alpha_{ij} ^p \cdot \textbf{h}_j),
\end{equation}
where $\textbf{h}_i ^p$ represents the learned embedding of node $i$ for meta-path $p$, and $\mathcal{N}_i ^p$ denotes the meta-path-based neighbors of node $v_i$. By obtaining the meta-path set $\{ p_0, p_1, \ldots, p_m \}$ through Formula (3), and applying node-level attention to node features, we derive $m$ sets of semantic-specific node embeddings, denoted as $\{ \mathcal{H}_{p_0},\ldots,\mathcal{H}_{p_m}\}$.

\subsection{Attribute-level Attention Mechanism}\label{sec:A2M}
Through conventional encoding, the resulting node embeddings learn the significance of different neighbors of nodes in each meta-path for the specific task at hand. However, they may fail to reflect the semantic importance of node attributes. A straightforward approach involves designing an attention mechanism capable of directly performing weighted summation or averaging on node attributes based on attention weights, subsequently aggregating them to the nodes to derive the final node representation:
\begin{equation}
    \beta_{ij} = att(x_i,x_j),  \forall j \in \mathcal{N}_i,
\end{equation}
\begin{equation}
    \textbf{h}_i = \sum_{j \in \mathcal{N}_i} \beta_{ij} \cdot x_j,
\end{equation} 
where $x_i$ represents the attributed information of node $v_i$, and $\mathcal{N}_i$ denotes the neighbors of node $v_i$. However, this simplistic attention mechanism treats all node types equally, disregarding the intricate relationships between different types of nodes in heterogeneous graphs. To address this, we propose a novel meta-path-based attribute-level attention mechanism to automatically learn the semantic importance of different attributes in meta-paths and fuse them for attributed community detection tasks. Taking $m$ groups of semantic-specific node embeddings learned from node-level attention as input, the learned weights of each meta-path $(\beta_{p_0},\ldots,\beta_{p_m})$ are calculated as follows:
\begin{equation}
    (\beta_{p_0},\ldots,\beta_{p_m}) = att_{sem}(\mathcal{H}_{p_0},\ldots,\mathcal{H}_{p_m}),
\end{equation} 
where $att_{sem}$ denotes the deep neural network performing the attribute-level attention. Initially, we average the importance of all semantic-specific node embeddings to determine the importance of each meta-path. The significance of each meta-path, denoted as $\mathscr{w}_{p_i}$, is computed as follows:
\begin{equation}
\mathscr{w}_{p_i}= \frac{1}{|\mathcal{V}|} \sum_{i \in \mathcal{V}} q^T \cdot tanh(\mathcal{W} \cdot \textbf{h}_i ^p+ b),
\end{equation}
where $\mathcal{W} \in \mathbb{R}^{d^{\prime} \times d}$ represents the weight matrix, $b \in \mathbb{R}^{d^{\prime} \times 1}$ denotes the bias vector, and $q \in \mathbb{R}^{d^{\prime} \times 1}$ signifies the semantic attention vector \cite{wang2019heterogeneous}. To this extent, we obtain the importance of each meta-path, which can help draw key attributes of communities. 

To learn the importance of different attributes in each meta-path and fuse the semantic similarity between attributes within the community, we propose to utilize the node similarity metric to update the importance of meta-paths. Recognizing the limitations of Euclidean distance in measuring node similarity in graph data due to the curse of dimensionality and differences in feature weight, we employ an attention-based similarity score \cite{chen2019graphflow}:
\begin{equation}
s_{ij} = (x_i \cdot u)^T \cdot x_j,
\end{equation} 
where $x_i$ and $x_j$ denote the attributed feature vectors of node $v_i$ and node $v_j$, respectively, and $u$ represents a non-negative trainable weight vector. Subsequently, we normalize the attention-based similarity scores to obtain the attribute coefficient $\gamma_{ij}$ as follows:
\begin{equation}
\gamma_{ij} = \frac{exp(s_{ij})}{\sum_{k \in \mathcal{N}_i ^p} exp(s_{ik})}.
\end{equation}

Intuitively, nodes with more similar attributes tend to exert greater influence on nodes within the target community. To adaptively adjust the relative importance of meta-path semantic and attributed semantic similarity, we introduce two learnable parameters $l_s ^p$ and $l_a ^p$, denoted as $q_s ^p$ and $q_a ^p$, respectively. They are formally expressed as follows:
\begin{equation}
    q_s ^p = \frac{exp(l_s ^p)}{exp(l_a ^p) + exp(l_s ^p)},
\end{equation}
\begin{equation}
    q_a ^p = \frac{exp(l_a ^p)}{exp(l_a ^p) + exp(l_s ^p)}.
\end{equation}

After that, we combine the meta-path coefficient $\mathscr{w}_{p_i}$ and the attribute coefficient $\gamma_{ij}$ to compute the attribute-level importance coefficient $\beta_{p_i}$:
\begin{equation}
    \beta_{p_i} = q_a ^p \cdot \sum_{i,j \in p_i} \gamma_{ij} + q_s ^p \cdot \mathscr{w}_{p_i},
\end{equation}

With the learned attribute-level importance coefficients, we fuse all semantic-specific embeddings to obtain the final embedding $\mathcal{H}$:
\begin{equation}
\mathcal{H}=\sum_{i=1} ^P \beta_{p_i} \cdot \mathcal{H}_{p_i}.
\end{equation}

This process further integrates the attribute-based semantic similarity, effectively capturing the nuanced relationships between attributes and nodes within the community.

However, the aforementioned process introduces a large number of trainable parameters to extract semantic information, necessitating the task of uncovering additional supervised signals to ensure training accuracy. To address this challenge, we integrate the concept of contrastive learning. Through self-supervised learning, we aim to unearth latent signals. Intuitively, the embedding of nodes within each community should exhibit similarity, ideally minimizing the distance between nodes within the community:
\begin{equation}
    L_{intra} = \sum_{c \in C} \sum_{i,j \in c} dist(\textbf{h}_i,\textbf{h}_j),
\end{equation} 
where $C$ represents the set of all communities. However, if we only consider similarities within communities, all nodes in the network may become similar, leading to the entire network being perceived as a single community. Therefore, it is crucial to ensure that node embeddings between communities are as dissimilar as possible. It can be formally expressed as:
\begin{equation}
    L_{inter} = \sum_{(c_1,c_2) \in E_c} \sum_{i \in c_1,j \in c_2} dist(\textbf{h}_i,\textbf{h}_j),
\end{equation} 
where $E_c$ denotes the set of edges between communities, and $c_1$, $c_2$ represent different communities. Inspired by recent work of contrastive learning~\cite{LCZ23,ZCL23}, which aims to learn effective representations by minimizing the distance between similar samples and maximizing the distance between dissimilar samples, we construct the objective function for attribute cohesiveness:
\begin{equation}
    L_A = r_1 L_{intra} - r_2 L_{inter},
\end{equation} 
where $r_1$ and $r_2$ serve as controlling parameters. This formulation encourages the embeddings of nodes within the same community to be similar while promoting dissimilarity between nodes from different communities.

\subsection{Community Membership Function}\label{sec:CMF}
A traditional approach to detecting communities relies on modularity optimization, typically employing greedy algorithms or constructing modularity matrices. High-quality communities exhibit high modularity, indicating dense connections within communities and sparse connections to nodes outside the community. However, directly capturing information from connections may lead to suboptimal results, as it overlooks the joint recognition of information from nodes, edges, and neighborhoods with special attention in the deep learning process. Deep neural network-based community detection frameworks embed complex structural relationships and minimize loss, such as cross-entropy, over all possible permutations $S_c$ of community labels:
\begin{equation}
    L = \mathop{\min}_{\pi \in S_c} - \sum_i \log o_{i, \pi(y_i)}.
\end{equation}
Here, the softmax function identifies conditional probabilities that a node $v_i$ belongs to the community $C_k$ ($o_{i,k}=p(y_i=c_k)$). Notably, these approaches primarily focus on microscopic pairwise connections, neglecting modularity, which can reveal the inherent community structure during training.

Drawing on recent research, we propose incorporating modularity into the training process to effectively capture the underlying community structure. However, the classical definition of modularity only emphasizes first-order proximity, which may oversimplify complex structures in real-world scenarios. To extend modularity to higher-order proximity, it requires redefinition:
\begin{equation}
\widetilde{Q}= \sum_{c_k \in C} \sum_{i,j} \varphi_{i,c_k} \varphi_{j,c_k} [\widetilde{A}_{ij}- \frac{\widetilde{k}_i \widetilde{k}_j}{2 \widetilde{M}}],
\end{equation}
where, $\varphi_{i,c_k}, \varphi_{j,c_k} \in [0,1]$, and $\sum_{c_k \in C} \varphi_{i,c_k}=\sum_{c_k \in C} \varphi_{j,c_k}=1$. Additionally, we encode node category labels into one-hot vectors, constructing the initial community membership matrix $M$ by integrating these vectors. Then, we concatenate $M$ horizontally with the feature matrix $X$ to form the updated matrix $X^{\prime}$. After training, the attributed community detection model saves the learned community membership information as community membership embedding $B$, corresponding to $\varphi_{i,c_k} \varphi_{j,c_k}$. We can rewrite $\widetilde{Q}$ in matrix terms:
\begin{equation}
\widetilde{Q}= \frac{1}{2 \widetilde{M}}tr(B^T \widetilde{P} B),
\end{equation} 
where $\widetilde{P}=\widetilde{A}_{ij}- \frac{\widetilde{k}_i \widetilde{k}_j}{2 \widetilde{M}}$. Based on the above, we design the CMF to guide embedding for preserving the inherent community structure. Modularity quantifies the disparity between the actual number of edges within a community and the anticipated number in a comparable network with randomly distributed edges. A higher modularity value indicates a stronger concentration of structural information within the community compared to random expectations. We formulate the CMF as a  modularity optimization problem:
\begin{equation}
L_M = - \widetilde{Q}.
\end{equation} 

\subsection{Training}
We prioritize the CMF as the main objective to obtain the division of communities on a global scale and further refine community members using the attribute cohesiveness function. The total loss is used for training as follows:            
\begin{equation}                                  
\mathcal{L} = L_M + \lambda \cdot L_A,         
\end{equation} where $\lambda$ is a controlling parameter that adjusts the impact of attribute cohesiveness. 
\section{Experiments}

In this section, we evaluate the effectiveness of our proposed HACD model on five real-world datasets by comparing it with seven state-of-the-art baseline methods.

\subsection{Experimental Setup}
\subsubsection{Datasets} 
We use five public benchmark datasets widely employed in community detection\cite{zhang2020commdgi,wu2022clare}: Cora, Citeseer, Amazon, Pubmed, and DBLP, all accessible from the SNAP website\footnote{https://snap.stanford.edu/data/index.html}. The distinct statistical properties of these different datasets make them suitable for reliably validating model performance. The statistics are summarized in Table~\ref{tab:data}.

\begin{table} 
  \caption{Benchmark graph datasets.}
  \label{tab:data}
  \begin{tabular}{ccccc}
    \toprule
    Dataset&\#Nodes&\#Edges&\#Features&\#Communities\\
    \midrule
    Cora & 2,708 & 5,429 & 1,433 & 7\\
    Citeseer & 3,327 & 4,732 & 3,703 & 6\\
    Amazon & 6,926 & 17,893 &599 &1,000\\
    Pubmed & 19,717 & 44,338 & 500 & 3\\
    DBLP & 37,020 & 149,501 &334 &1,000\\
  \bottomrule
\end{tabular}
\end{table}

\subsubsection{Baseline algorithms}
To demonstrate the effectiveness of HACD, we compare it with several state-of-the-art methods:
\begin{itemize} 
\item \textbf{GCN}\cite{kipf2016semi}: A fundamental graph representation learning model that operates directly on graph-structured data.
\item \textbf{GAT}\cite{velivckovic2017graph}: A model that leverages masked self-attentional layers to assign different weights to different nodes in a neighborhood.
\item \textbf{AnECI}\cite{liu2022robust}: A framework for learning community information-based attributed network embedding by reconstructing higher-order proximity.
\item \textbf{CDE}\cite{li2018community}: A method that encodes potential community membership information based on nonnegative matrix factorization (NMF) optimization.
\item \textbf{DANMF}\cite{ye2018deep}:  A deep autoencoder-like nonnegative matrix factorization model for community detection.
\item \textbf{DAEGC}\cite{wang2019attributed}: A goal-directed graph clustering approach that employs an attention network to encode the importance of neighboring nodes and reconstructs the graph structure by training a decoder.
\item \textbf{CommDGI}\cite{zhang2020commdgi}: A community detection-oriented graph neural network that uses a mutual information mechanism to capture neighborhood and community information.
\end{itemize}

\subsubsection{Evaluation Metrics and Parameter Settings}
We use five widely adopted metrics to measure the performance of the methods: accuracy (ACC), F1-score (F1), normalized mutual information (NMI), adjusted rand index (ARI), and modularity. A better model should exhibit higher values across all metrics.

\begin{table*}
    \caption{The performance of different ACD methods. The best results are boldfaced, and the second-best results are underlined.}
    \begin{tabular}{cccccccccc}
      \toprule
      \textbf{Dataset} & \textbf{Metric} &GCN &GAT &AnECI &CDE &DANMF &DAEGC &CommDGI &HACD\\
      \midrule
      \multirow{5}{5em}{\textbf{Cora}} &ACC &0.3383 &0.3298 &\underline{0.3567} &0.2563 &0.2010 &0.3051 &0.2758 &\textbf{0.5916}  \\
      & NMI &0.1142 &0.1556 &0.1308 &0.1654 &0.1021 &0.1796 &\underline{0.1919} &\textbf{0.4030}  \\
      & ARI &0.1055 &0.1032 &0.1248 &0.1402 &0.0991 &0.1248 &\underline{0.1541} &\textbf{0.3260}  \\
      & F1 &0.2071 &0.2450 &0.2242 &\underline{0.2658} &0.2062 &0.2348 &0.2484 &\textbf{0.4803} \\
      & Modularity &0.0990 &0.1905 &\underline{0.7160} &0.3870 &0.3974 &0.5392 &0.5797 &\textbf{0.7364}  \\
      \midrule
      \multirow{5}{5em}{\textbf{Citeseer}} &ACC &0.2852 &\underline{0.2985} &0.2453 &0.2579 &0.2271 &0.2647 &0.2549 &\textbf{0.4740}  \\
      & NMI &0.1853 &0.1779 &0.1280 &0.1434 &0.1678 &0.2013 &\underline{0.2252} &\textbf{0.3130}  \\
      & ARI &0.1569 &0.1253 &0.1168 &0.1228 &0.1704 &0.2063 &\underline{0.2204} &\textbf{0.2608}  \\
      & F1 &0.2153 &0.2439 &0.1572 &0.2365 &0.2056 &0.2340 &\underline{0.2464} &\textbf{0.4534} \\
      & Modularity &0.1853 &0.2870 &\underline{0.8137} &0.2309 &0.2195 &0.2623 &0.2136 &\textbf{0.8388}  \\
      \midrule
      \multirow{5}{5em}{\textbf{Amazon}} &ACC &0.1391 &0.1368 &0.1729 &0.1213 &0.0934 &0.1176 &\underline{0.2254} &\textbf{0.2828}  \\
      & NMI &0.3195 &0.2994 &0.3107 &0.2651 &0.1677 &0.2165 &\underline{0.3630} &\textbf{0.6056}  \\
      & ARI &0.1142 &0.1097 &0.1436 &0.0817 &0.0603 &0.1039 &\underline{0.1589} &\textbf{0.1684}  \\
      & F1 &0.1636 &0.1566 &0.2949 &0.1419 &0.1363 &0.1016 &\textbf{0.3846} &\underline{0.3252} \\
      & Modularity &0.1263 &0.1897 &\textbf{0.9783} &0.2174 &0.2409 &0.2630 &0.3415 &\underline{0.6792}  \\
      \midrule
      \multirow{5}{5em}{\textbf{Pubmed}} &ACC &0.4625 &\underline{0.4823} &0.3995 &0.1293 &0.2108 &0.4455 &0.4181 &\textbf{0.7034}  \\
      & NMI &0.1368 &0.2028 &0.1141 &0.0583 &0.1057 &0.1632 &\underline{0.3276} &\textbf{0.4346}  \\
      & ARI &0.1464 &0.1902 &0.1493 &0.0511 &0.0489 &0.2403 &\underline{0.2597} &\textbf{0.3988}  \\
      & F1 &0.3926 &\underline{0.4348} &0.3417 &0.1142 &0.1067 &0.4333 &0.3889 &\textbf{0.6103} \\
      & Modularity &0.2588 &0.2351 &\underline{0.6071} &0.3793 &0.3275 &0.3844 &0.4962 &\textbf{0.6529}  \\
      \midrule
      \multirow{5}{5em}{\textbf{DBLP}} &ACC &0.1016 &0.1173 &0.1337 &0.0034 &0.0925 &0.0094 &\underline{0.2365} &\textbf{0.4274}  \\
      & NMI &0.1430 &0.1379 &0.2903 &0.0036 &0.1022 &0.0014 &\underline{0.3104} &\textbf{0.3185}  \\
      & ARI &0.0414 &0.0752 &0.0024 &0.0057 &0.0041 &0.0002 &\underline{0.1119} &\textbf{0.2041}  \\
      & F1 &0.1886 &0.1560 &0.0941 &0.0323 &0.1108 &\underline{0.3385} &0.2057 &\textbf{0.3492} \\
      & Modularity &0.1172 &0.1641 &\textbf{0.8106} &0.2137 &0.2082 &0.1192 &0.4472 &\underline{0.7515}  \\
      \bottomrule
    \end{tabular}
    \label{tab:q1}
\end{table*}

\subsubsection{Parameter Settings}
We train our model for 400 iterations and maintain a fixed size of 32 for the embeddings. We use Adam to optimize the parameters with a default learning rate of 0.01 and a default weight decay of 0.2. For the baseline algorithms, we meticulously set all hyper-parameters according to the scope outlined in their original papers and tune them on every datasets.

\subsection{Experiment Results}
\subsubsection{Overall Performance} 
We compare the performance of our HACD model with seven state-of-the-art community detection methods on five real-world datasets, as shown in Table~\ref{tab:q1}. We can make the following key observations:
\begin{itemize}
    \item Our proposed HACD framework achieves noticeable improvements across nearly all datasets. Among them, compared to the baselines with the best results, HACD respectively achieves the highest improvement of 23.49\%, 24.26\%, 17.19\%, 21.45\%, and 4.58\% in five evaluation indicators, demonstrating its effectiveness. Notably, HACD achieves significant performance gains on the Pubmed and DBLP datasets. This not only validates our method but also highlights HACD’s capability to detect communities in large-scale datasets.
    \item GNN-based methods generally outperform CDE and DANMF, due to the excellent performance of GNN in mining node attribute information. 
    However, since GNN-based baselines only incorporate intuitive attribute information without delving into the semantic similarity between attributes, they cannot fully utilize the information within the attributes.

    \item CDE, CommDGI, and AnECI achieve good results in almost all evaluation metrics. Different from capturing structural performance by encoding the pairwise connections of nodes, they encode latent community membership information, demonstrating the efficiency of leveraging inherent community information. However, while CDE, CommDGI, and AnECI encode membership information, they overlook higher-level mesoscopic structural constraints and global structural patterns, resulting in suboptimal performance.
    
\end{itemize}
\begin{table*}
\caption{F1-score (in \%) and time cost (in seconds) of baselines and HACD on all datasets.}
\centering
    \begin{tabular}{c||cc|cc|cc|cc|cc}
      \toprule
      \multirow{2}{5em}{\textbf{Methods}} & \multicolumn{2}{c}{\textbf{Cora}} & \multicolumn{2}{c}{\textbf{Citeseer}} & \multicolumn{2}{c}{\textbf{Amazon}} & \multicolumn{2}{c}{\textbf{Pubmed}} & \multicolumn{2}{c}{\textbf{DBLP}} \\
       &F1 &Time &F1 &Time &F1 &Time &F1 &Time &F1 &Time \\
      \midrule
      \multirow{1}{5em}{\textbf{GCN}} &20.71 &18.71 &21.53 &30.80 &16.36 &139.82 &39.26 &279.67 &18.86 &998.06 \\
      \multirow{1}{5em}{\textbf{GAT}} &24.50 &18.43 &24.39 &27.71 &15.66 &155.69 &43.48 &279.26 &15.60 &1047.15 \\
      \multirow{1}{5em}{\textbf{AnECI}} &22.42 &15.54 &15.72 &28.33 &29.49 &138.57 &34.17 &301.08 &9.41 &1061.60 \\
      \multirow{1}{5em}{\textbf{CDE}} &26.58 &20.07 &23.65 &23.71 &14.19 &120.01 &11.42 &283.51 &3.23 &1003.29 \\
      \multirow{1}{5em}{\textbf{DANMF}} &20.62 &27.19 &20.56 &28.98 &13.63 &127.93 &10.67 &303.59 &11.08 &1125.97 \\
      \multirow{1}{5em}{\textbf{DAEGC}} &23.48 &19.43 &23.40 &29.72 &10.16 &71.92 &43.33 &268.97 &33.85 &979.15 \\
      \multirow{1}{5em}{\textbf{CommDGI}} &22.39 &35.05 &21.06 &45.01 &30.45 &1596.96 &36.31 &411.93 &19.36 &2127.11 \\
      \multirow{1}{5em}{\textbf{HACD}} &\textbf{41.85} &20.88 &\textbf{27.05} &37.11 &\textbf{30.53} &140.90 &\textbf{56.31} &293.14 &\textbf{34.01} &1083.19\\
      \bottomrule
    \end{tabular}
    \label{tab:time}
\end{table*}
\subsubsection{Attribute Information} 
\begin{figure}
  \includegraphics[width=0.9\linewidth]{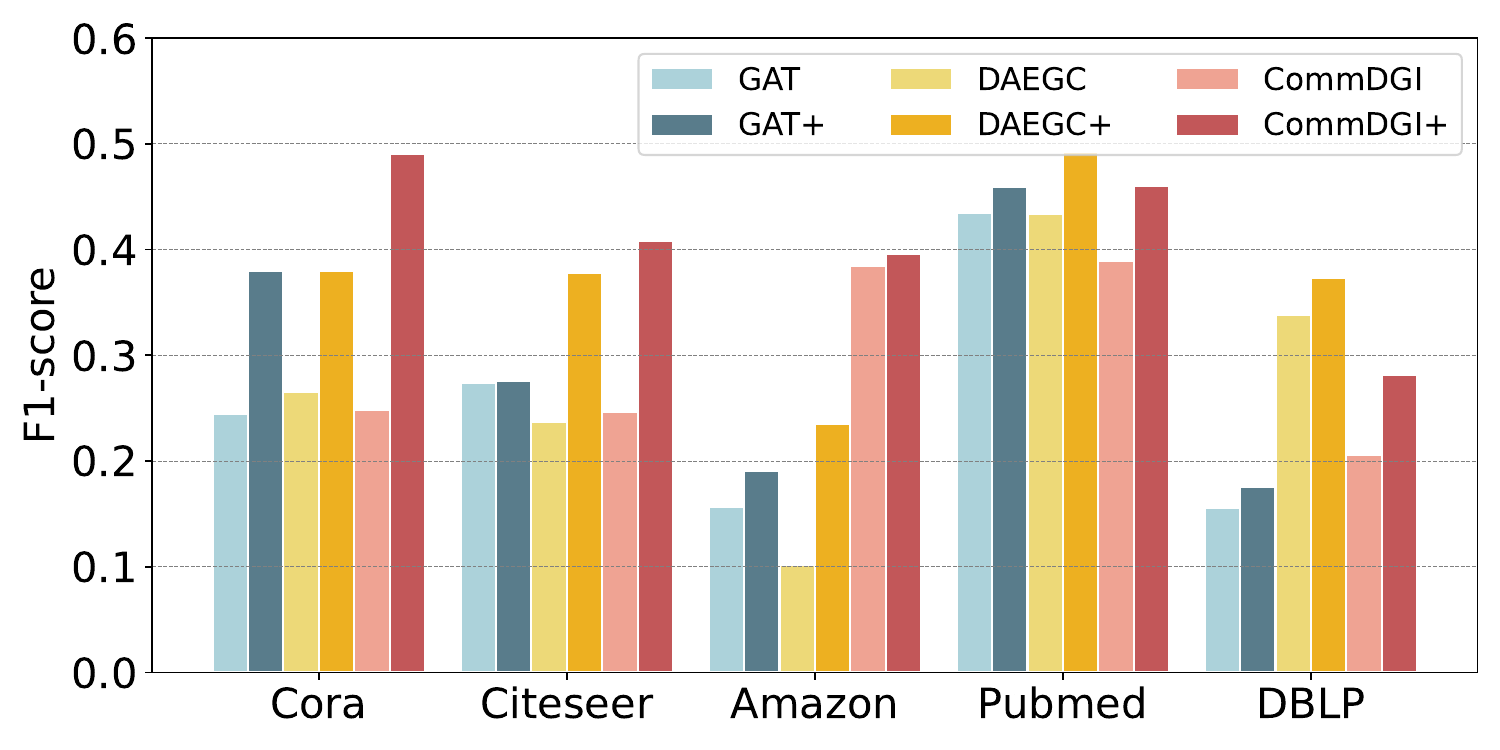}
  \caption{The impact of the original graph structure and the updated graph structure for model performance.}
  \label{fig:updated}
\end{figure}

Instead of using the original attributed graph structure directly, our model pioneers a new approach. We treat node attributes as another type of node, transforming real-world attributed graphs into a heterogeneous graph structure. We then apply this updated graph structure to baseline models such as GAT, DAEGC, and CommDGI, which also consider attribute information. Figure~\ref{fig:updated} shows the performance comparison between the models using the original graph structure and the corresponding improved models. It is obvious that the improved models universally outperform the original models, demonstrating that using the updated graph structure as input allows each model to encode attribute information at a higher level of granularity, resulting in improved performance. This proves that our enhanced graph structure can unlock the potential of attribute information.

\subsubsection{Efficiency}
We evaluate the efficiency of HACD by directly comparing the total running time with all baselines. In this evaluation, all models run 600 epochs as well as other parameters for baselines are set following their original papers. Table~\ref{tab:time} illustrates the performance (F1 score) and running time (seconds). We can observe that the running time of HACD is consistently competitive. HACD is always faster than CommDGI, which also considers dual information of attributes and community. Even on large-scale datasets Pubmed and DBLP, the running time of HACD is still within a reasonable range. 

\subsubsection{Parameter Discussions} 
\begin{figure}
  \subfigure[]{\label{fig:epoch}
    \includegraphics[width=0.45\linewidth]{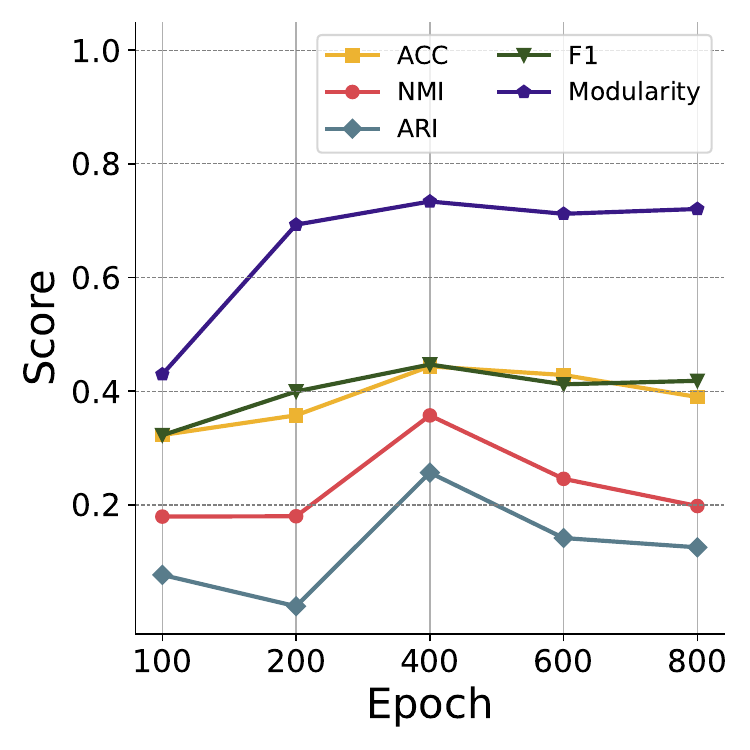}}\subfigure[]{\label{fig:embedding}
    \includegraphics[width=0.45\linewidth]{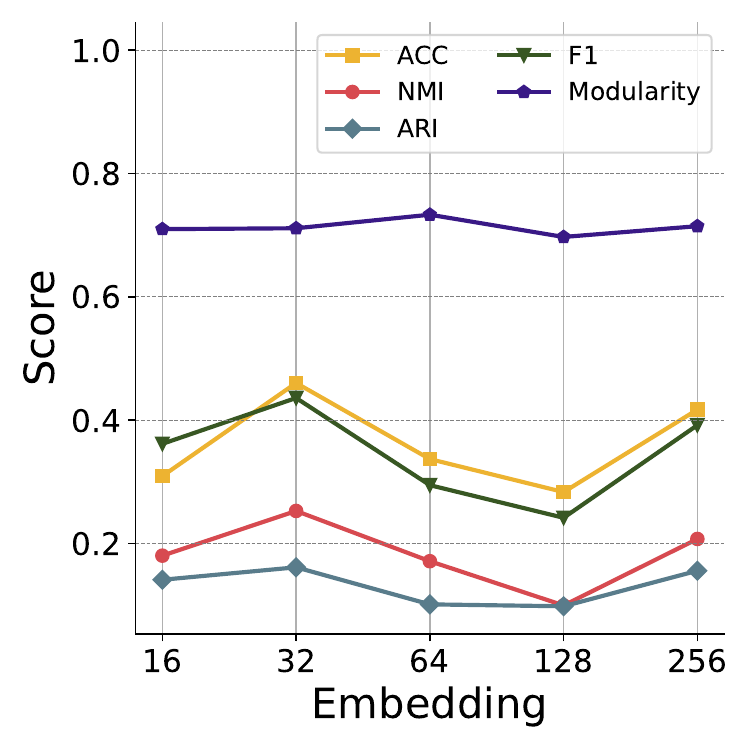}}
  \caption{The impact of parameters on HACD on the Cora dataset.}
  \label{fig:B}
\end{figure}

We vary the training epoch and the dimension of embedding to explore the parameter settings of our model. It can be observed from Figure~\ref{fig:epoch} and ~\ref{fig:embedding} that: (i) with the increase of parameter values, the trend initially rises and then declines. Because when training for fewer epochs or embedding sizes, HACD fails to sufficiently learn the data features but training for too many epochs or embedding sizes leads to overfitting. (ii) Due to the influence of the training epoch, modularity gradually increases and then maintains a stable level. 
\subsubsection{Robustness and Scalability}
\begin{figure}
  \subfigure[]{\label{fig:robustness}
    \includegraphics[width=0.45\linewidth]{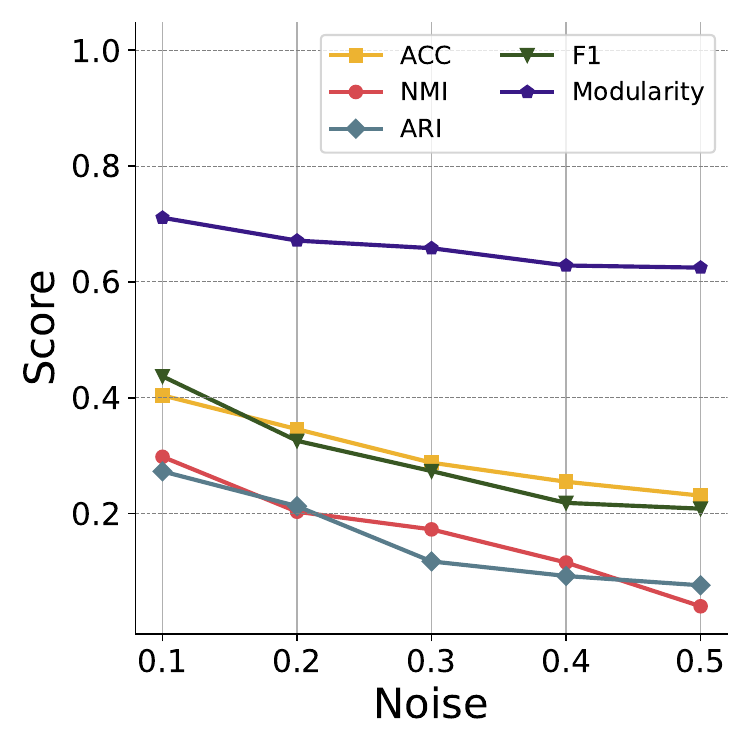}}\subfigure[]{\label{fig:scalability}
    \includegraphics[width=0.45\linewidth]{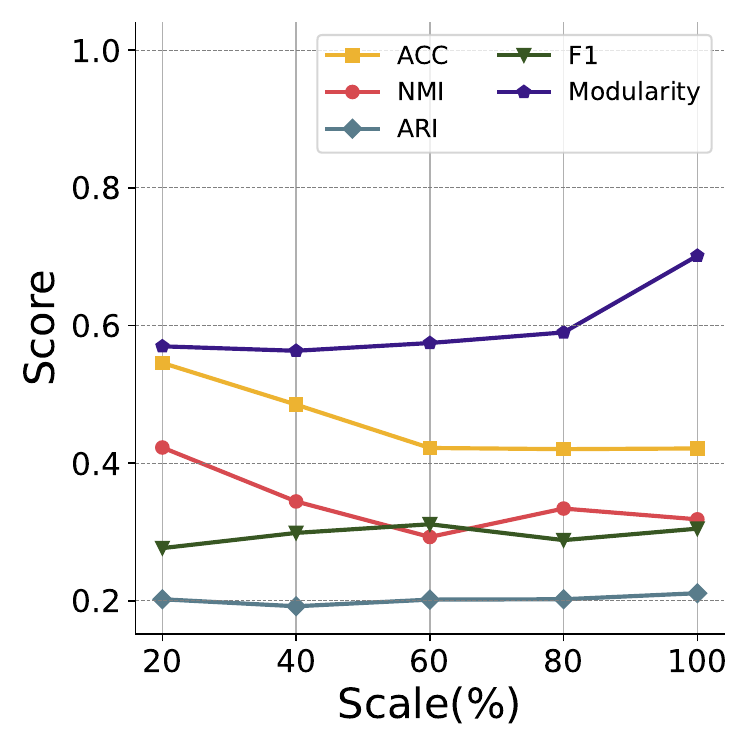}}
    \subfigure[]{\label{fig:scala_time}
    \includegraphics[width=0.8\linewidth]{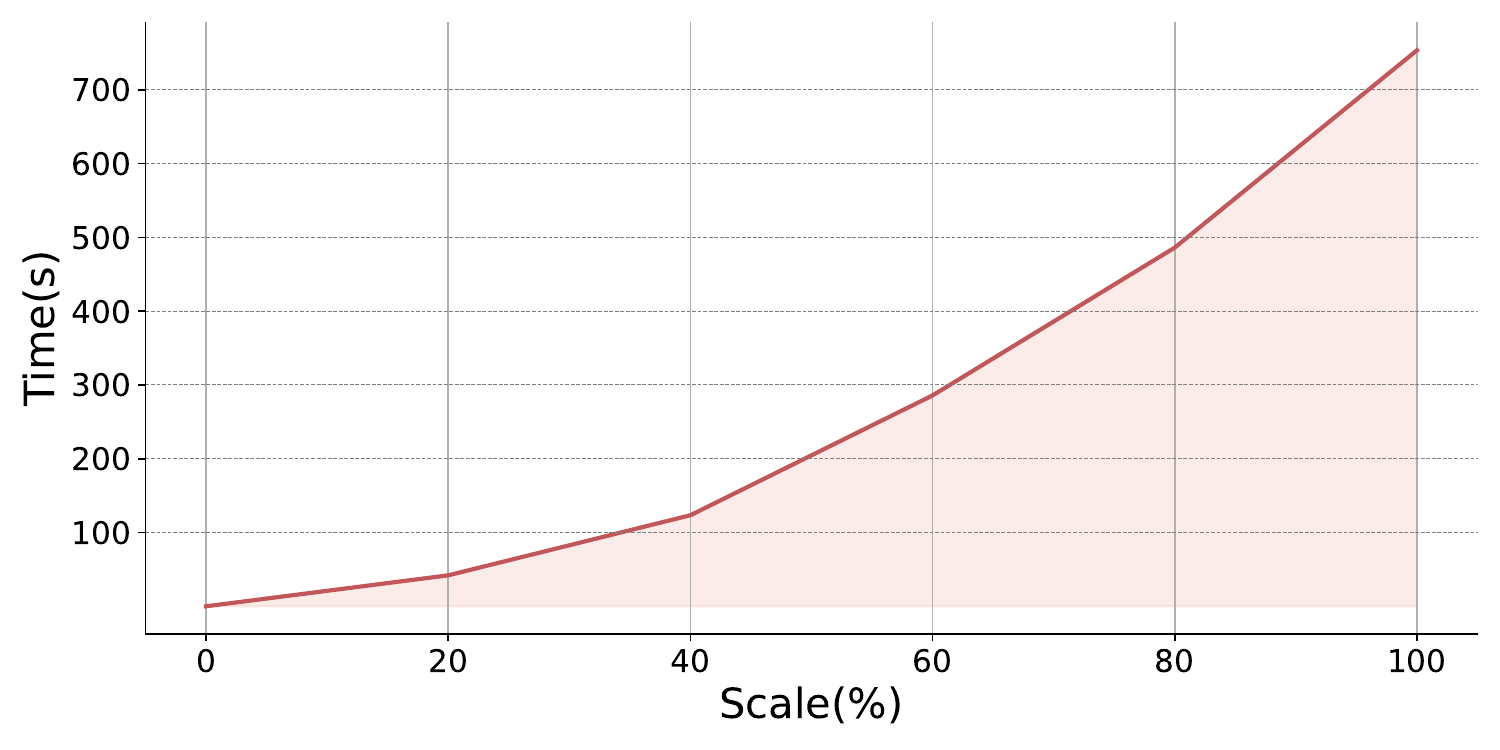}}
  \caption{The robustness and scalability of  HACD on the DBLP dataset.}
  \label{fig:C}
\end{figure}

We now discuss the robustness and scalability of our proposed model on the DBLP dataset. We add Gaussian noise to the network and verified the robustness of HACD by changing the range of data fluctuations. Figure~\ref{fig:robustness} shows the changes in the evaluation metrics. As the range of noise distribution expands, the modularity only decreases by about 5\% and then tends to stabilize, with little impact from noise variation. Although other evaluation metrics are more affected by noise, the decrease is still within a controllable range. Because HACD not only considers the information between node attributes at the microscopic level, but also takes into account the structural patterns at the mesoscopic level, exhibiting excellent robustness.

Figure~\ref{fig:scalability} and ~\ref{fig:scala_time} represent the evaluation metrics and running time for dataset of different scales, respectively. We find that: (i) the evaluation metrics are minimally affected by the scale of the dataset and remain relatively stable. (ii) The running time of HACD has not increased sharply due to the expansion of data scale, on the contrary, its growth rate is slow. These fully demonstrate the scalability of our model.

\subsubsection{Ablation Study} 
To validate the effectiveness of each part of HACD, we perform ablation experiments. The NMI results are shown in Tables~\ref{tab:NMI}. Due to space limitations, we omit the results of ACC, F1 and ARI, which show similar trends to NMI. A2M and CMF denote HACD utilizing only the attribute-level attention module and the community attribution function module, respectively

From Table~\ref{tab:NMI}, it can be observed that both A2M and CMF have improved clustering metric results to varying degrees, demonstrating their effectiveness. Specifically, A2M leverages meta-paths to capture diverse semantic information and further explores the semantic similarity between node attributes along each meta-path. This tendency leads nodes with more similar attribute semantics to be grouped into the same community, achieving finer-grained node classification within the network and yielding better clustering metrics. On the other hand, CMF utilizes node labels as the basis for initial community assignment, improving clustering metric results. However, CMF primarily focuses on optimizing global community coherence, resulting in inferior performance compared to A2M.

For the community structure evaluation metric Modularity, the results are presented in Table~\ref{tab:Modularity}. A2M focuses on the semantic similarity between node attributes, optimizing the community membership composition from an individual node perspective, thus affecting the overall network cohesion and enhancing the tightness of communities to a certain extent. On the other hand, CMF not only encodes implicit community affiliation information but also leverages high-order modularity information to guide model training, dividing communities from a global perspective. Therefore, the improvement in Modularity is more significant with CMF.

HACD integrates A2M and CMF. Specifically, HACD ensures the homogeneity of nodes within communities by leveraging semantic similarity between node attributes, achieving attribute cohesiveness within communities. Moreover, it utilizes inherent community information to enforce global structural patterns, thereby achieving the structural cohesiveness of communities in the whole network. HACD significantly outperforms A2M and CMF in all metrics, demonstrating that the combined implementation of these aspects comprehensively enhances the quality of detected communities.

\begin{table}
  \caption{Ablation study on NMI score.}
  \label{tab:NMI}
  \begin{tabular}{cccccc}
    \toprule
     &\textbf{Cora}&\textbf{Citeseer}&\textbf{Amazon}&\textbf{Pubmed}&\textbf{DBLP}\\
    \midrule
    \textbf{A2M} & 0.3576 & 0.2074 & 0.3923 &0.3601 &0.2339 \\
    \textbf{CMF} & 0.1741 & 0.1143 & 0.5712 &0.1039 &0.1097 \\
    \textbf{HACD} & \textbf{0.4030} & \textbf{0.3130} & \textbf{0.6056} &\textbf{0.4346} &\textbf{0.3185} \\
  \bottomrule
\end{tabular}
\end{table}
\begin{table}
  \caption{Ablation study on Modularity score.}
  \label{tab:Modularity}
  \begin{tabular}{cccccc}
    \toprule
     &\textbf{Cora}&\textbf{Citeseer}&\textbf{Amazon}&\textbf{Pubmed}&\textbf{DBLP}\\
    \midrule
    \textbf{A2M} & 0.1157 & 0.1817 & 0.2384 &0.2640 &0.2026 \\
    \textbf{CMF} & 0.6930 & 0.7505 & 0.8174 &0.6468 &0.7053 \\
    \textbf{HACD} & \textbf{0.7364} & \textbf{0.8388} & \textbf{0.6792} &\textbf{0.6529} &\textbf{0.7515} \\
  \bottomrule
\end{tabular}
\end{table}

\section{Related Work}

\subsection{Community Detection}

Community detection \cite{jin2021survey,su2022comprehensive,Ren2020overlapping} is commonly defined as the process of partitioning graph nodes into multiple groups and widely applied in various real-world applications, such as recommendation system\cite{satuluri2020simclusters} and anomaly detection \cite{yu2023group}. In recent years, graph neural networks (GNNs) have proven effective in various graph data mining tasks and exhibit strong capabilities in community detection \cite{luo2021detecting,shan2024kpi}. CP-GNN \cite{luo2021detecting} uses a context path-based GNN to detect communities in heterogeneous graphs, and KPI-HGNN \cite{shan2024kpi} designs a community detection algorithm based on heterogeneous graph neural network. 


Attributed graphs integrate attributes into the graph structure, resulting in a richer network representation \cite{sun2020nec}. Attributed community detection \cite{zhu2020structure, 2024Evolutionary} aims to find densely connected communities with homogeneous attributes by leveraging both topological and attribute information. Method like CDE \cite{li2018community} formulates the problem as a NMF optimization task, while ACDM \cite{cheng2023significant} constructs an attributed k-NN layer to extract common node representations. Recently, COD\cite{Niu2024Discovering} devises a local hierarchical reclustering method to identify the largest community, which takes into account the query attribute. 

Despite the widespread use of GNNs in non-attributed community detection \cite{ma2022convex} and graph clustering \cite{Daneshfar2024clustering}, their application to attributed community detection remains underdeveloped. Moreover, existing ACD methods often overlook the inherent community structures and encode node attributes directly, neglecting the semantic similarities between attributes within real communities. Our model effectively addresses these two issues by integrating A2M and CMF at the same time.

\section{Conclusion}
In this paper, we study the problem of attributed community detection from a new heterostructure perspective. We propose HACD, a model that ensures both attribute cohesiveness and structure cohesiveness in detected communities. Specifically, we construct attributed networks into a heterogeneous graph structure. We then use A2M to capture attribute semantic similarity and reveal the latent relationships between nodes in the network. Finally, CMF addresses sensitivity issues and enhances robustness by optimizing the community structure. Extensive experiments on real-world datasets demonstrate that our HACD model effectively discovers communities in attributed networks and significantly outperforms all baseline methods. While our model has demonstrated promising results, there remain opportunities to enhance its interpretability and generalization capabilities. In future work, we will explore alternative graph structure optimization techniques to further strengthen these aspects and investigate the novel insights that may emerge from the interplay between large language models for graphs and community detection.

\bibliographystyle{acm}

\bibliography{HACD-wsdm}

\end{document}